\documentclass[sigconf]{acmart}

\copyrightyear{2019}
\acmYear{2019}
\setcopyright{iw3c2w3}
\acmConference[WWW '19]{Proceedings of the 2019 World Wide Web Conference}{May 13--17,
2019}{San Francisco, CA, USA}
\acmBooktitle{Proceedings of the 2019 World Wide Web Conference (WWW '19), May 13--17, 2019,
San Francisco, CA, USA}
\acmPrice{}
\acmDOI{10.1145/3308558.3313696}
\acmISBN{978-1-4503-6674-8/19/05}

\usepackage{balance}
\usepackage{booktabs} 
\usepackage{graphicx}
\usepackage[utf8x]{inputenc} 
\usepackage{tikz}
\usepackage{subfig}

\begin{document}

\title{Keyphrase Extraction from Disaster-related Tweets}


\author{Jishnu Ray Chowdhury}
\affiliation{
  \institution{Kansas State University}
  \city{Manhattan}
  \state{KS}
}
\email{jishnurayc@ksu.edu}

\author{Cornelia Caragea}
\affiliation{
  \institution{University of Illinois at Chicago}
  \city{Chicago}
  \state{IL}
}
\email{cornelia@uic.edu}

\author{Doina Caragea}
\affiliation{%
  \institution{Kansas State University}
  \city{Manhattan}
  \state{KS}
  }
\email{dcaragea@ksu.edu}

\begin{abstract}
While keyphrase extraction has received considerable attention in recent years, relatively few studies exist on extracting keyphrases from social media platforms such as Twitter, and even fewer for extracting disaster-related keyphrases from such sources. During a disaster, keyphrases can be extremely useful for filtering relevant tweets that can enhance situational awareness. Previously, joint training of two different layers of a stacked Recurrent Neural Network for keyword discovery and keyphrase extraction had been shown to be effective in extracting keyphrases from general Twitter data. We improve the model's performance on both general Twitter data and disaster-related Twitter data by incorporating contextual word embeddings, POS-tags, phonetics, and phonological features. Moreover, we discuss the shortcomings of the often used F1-measure for evaluating the quality of predicted keyphrases with respect to the ground truth annotations. Instead of the F1-measure, we propose the use of embedding-based metrics to better capture the correctness of the predicted keyphrases. In addition, we also present a novel extension of an embedding-based metric. The extension allows one to better control the penalty for the difference in the number of ground-truth and predicted keyphrases. 

\end{abstract}

%
%



\maketitle
\section{Introduction}
Keyphrase extraction is the task of automatically extracting words or phrases from a text, which concisely represent the essence of the text.
Keyphrases can be used for multiple tasks including text summarization, classification, and opinion mining. They can also be highly valuable for retrieving relevant documents when searching for specific information. In particular, keyphrases can be used to organize and retrieve useful information from Twitter.

Twitter is an online social media platform, where people communicate via short, terse messages (``tweets''). Due to the ease of posting a tweet, and the popularity of the platform, Twitter can provide up-to-date news around the world, faster than standard media outlets. Thus, in recent years, Twitter has become an important source of real-time information \cite{Tran:2016,Wen:2017}.
Specifically, during crisis situations, Twitter has revolutionized the way in which affected populations communicate with response organizations \cite{bruno2011tweet,cooper2015twitter}. On one hand, the affected populations can follow and spread the latest updates from government and response organizations on Twitter. On the other hand, affected individuals can post information that can be used to enhance situational awareness and, in some cases, save lives. Emergency hotlines, such as 911, could be overwhelmed by the large number of calls received during major disasters, and as a consequence, individuals often turn to social media, including Twitter, to request assistance
\cite{Rhodan:2017,MacMillan:2017,Frej:2018,Lapin:2018}. For example, a North Carolina family was rescued during Hurricane Florence after posting the following message on Twitter: \emph{`If anybody could help… our car is under water and so is our house. Stuck in the attic. Phone is about to die. Please send help'} \cite{Lapin:2018}.

While social media can help enhance situational awareness, facilitate rescue operations, and ultimately save lives, its value is highly unexploited by response teams, in part due to the lack of tools that can help filter relevant, informative, and actionable information posted during disasters \cite{Villegas:2018}. 
To address this limitation, the annotation of tweets with informative keyprhases can enable filtering of actionable information by the disaster management and response teams, in real time. In fact, in Twitter, hashtags assigned directly by the users posting the tweets are sometimes used to annotate tweets and can be seen as keyprases useful for filtering. However, a vast majority of tweets do not have any hashtags associated with them, as illustrated by a dataset we crawled during several crisis events that happened in the fall of 2017. Thus, to speed up the search and retrieval of potentially actionable disaster tweets posted by affected individuals, there is an urgent need for methods that can automatically extract keyphrases from tweets without primarily relying on hashtags. The extracted keyphrases can potentially be recommended to users as they are typing their messages, and thus a smaller number of relevant tweets would be retrieved by systems that facilitate filtering of information for the response teams. 

Keyphrase extraction from tweets, however, is a non-trivial task due to the tweets' informal nature, noisiness 
and short length (currently, tweets have a limit of 280 characters). These factors limit the utility of many of the existing unsupervised keyword extraction techniques, which are designed to work on longer text using term-frequencies, word co-occurrences, and clustering methods. 
Recently, 
researchers started to explore keyphrase extraction 
from single tweets using deep-learning models. For example, \citeauthor{zhang2016keyphrase} \shortcite{zhang2016keyphrase} proposed a model based on joint-layer Recurrent Neural Networks (RNNs) and achieved an F1-score of $80.97\%$ when used to extract keyphrases from the general Twitter. However, this RNN-based model has not been extended to extract keyphrases from tweets posted in a specific context, e.g., during a disaster.  In fact, there is a distinctive lack of works for extracting keyphrases from disaster tweets. This task is especially challenging as the vocabulary used during a disaster is focused and limited, as compared to the vocabulary in a general collection of tweets. Furthermore, the keyphrases extracted should reflect information related to the event, people involved and their locations, the type of assistance needed, the type of situational awareness provided, etc.    
Thus, our research agenda is designed to answer the following questions:
\setlength{\parskip}{0em}
\begin{enumerate}
\item {\em Can we enhance the existing model to improve its performance on both disaster tweets and general tweets?} 
\item {\em How do the joint-layer-Recurrent Neural Network based models trained on general tweets perform on disaster tweets?}
\item {\em How well do the joint-layer-Recurrent Neural Network models perform when trained and tested on disaster tweet data?}
\end{enumerate}
Regarding question (1), we discovered that adding contextual word information, phonetics and phonological features, along with part of speech (POS) tag information, tends to increase the performance of the model on both the general domain and the disaster domain, when training the model on data from the relevant domain.
Regarding question (2), it is expected that the model will not perform very well if trained on a general-domain Twitter data and tested on a specific-domain data (e.g., disaster data). However, considering that the general training data and the specific disaster test data are both collected from Twitter, it is still interesting to empirically find the degree to which the RNN-based model trained on general Twitter data can or cannot generalize to disaster data. Finally, regarding question (3), while it is expected that the model will have better performance when trained and tested on specific disaster tweet data, it is interesting to understand how significant the improvement is and what additional information helps the most.\setlength{\parskip}{0em}

While investigating the above research questions, we found that the exact match F1-based evaluation is not ideal for determining the quality of the extracted keyphrases.  As an alternative to the exact match  F1 metric, we explored embedding-based similarity metrics \cite{foltz1998measurement,rus2012comparison, mitchell2008vector,forgues2014bootstrapping}, and proposed a novel extension to symmetric greedy embedding based similarity scoring. The extension allows for more fine-grained control over the penalty for the difference in the number of ground-truth and predicted keyphrases. 

We describe our extensions to the original RNN-based model for keyphrase extraction \cite{zhang2016keyphrase} in Section 3. In Section 4, we present the details of our datasets. We describe the experimental setting in Section 5.1, and the details of our embedding-based evaluation metric in Section 5.2. Finally, we present and discuss our results in Section 5.3 and conclude the paper in Section 6. 

\section{Related Work}

Keyphrase extraction can be supervised or unsupervised. Supervised approaches treat keyphrase extraction as a classification problem. In supervised approaches, a model is trained to learn to classify keyphrases from training data that is annotated with keyphrases \cite{CarageaBGG14,Frank:1999:DKE:646307.687591, Turney1999LearningTE, Turney:2000:LAK:593957.593993,Hulth:2003:IAK:1119355.1119383,witten2006thesaurus,Lopez:2010:HAK:1859664.1859719}.  Many unsupervised keyphrase extraction techniques had also been previously proposed \cite{sparck1972statistical,Salton:1986:IMI:576628,mihalcea2004textrank,liu2009clustering,rose2010automatic,litvak2011degext,GollapalliC14,FlorescuC17}. They usually extract candidate keyphrases and rank them based on term frequencies, word co-occurrences, and other similar features. However, the length of tweets is usually too short for most of these techniques to be applicable. Currently, tweets are limited to a length of 280 characters, with the average length of the tweets in a collection being even smaller. Furthermore, keyphrases will rarely occur more than once in a single tweet. These issues along with the noisiness of tweets can make keyphrase extraction from social media platforms such as Twitter a difficult task. 

Some of the previous works on keyphrase extraction from Twitter \cite{zhao2011topical,bellaachia2012ne} take the approach of inferring keyphrases from multiple tweets. Thus, they can only extract some topical information and trends that are common to multiple tweets. At the level of multiple tweets, some of the word-count based approaches for keyphrase extraction can be still applicable. However, our target is to extract keyphrases from single tweets. 
Along this line, the approaches proposed by 
\citeauthor{marujo2015automatic} \shortcite{marujo2015automatic} and \citeauthor{zhang2016keyphrase} \shortcite{zhang2016keyphrase} work for single tweets. \citeauthor{marujo2015automatic}
\shortcite{marujo2015automatic}
showed that word embeddings in a system such as MAUI \cite{medelyan2009human} performs better than the TF-IDF baseline \cite{sparck1972statistical}. Furthermore, the Joint-Layer-RNN model proposed by \citeauthor{zhang2016keyphrase} \shortcite{zhang2016keyphrase} was shown to be even better than the model proposed by \citeauthor{marujo2015automatic} \shortcite{marujo2015automatic}. The joint-layer RNN model can predict keyphrases of any arbitrary length, based on single tweets, without any significant pre-processing or hand-engineered features. Thus, we chose to focus on this model in our study. 

Regarding retrieval and filtering of tweets during disasters, there are many previous works related to processing social media during emergency situations \cite{imran2015processing}, extracting information from Twitter to enhance situational awareness  \cite{verma2011natural,vieweg2010microblogging,kumar2011tweettracker}, and classifying or clustering disaster-relevant tweets \cite{ashktorab2014tweedr,zhang2016semi,stowe2016identifying}, among others. However, to the best of our knowledge, no existing works have focused specifically on the extraction of keyphrases from disaster tweets, a task that we focus on in this study. 

Finally, in terms of evaluation, there are previous works related to embedding-based similarity metrics \cite{foltz1998measurement,rus2012comparison, mitchell2008vector,forgues2014bootstrapping}, which we also investigate. These metrics have been used for paraphrase detection, assessment of student input \cite{rus2012comparison}, and other similar tasks including assessment of the quality of generated dialogues \cite{liu2016not, serban2017hierarchical} by conversational models. We show that embedding-based similarity metrics can also be used to compare the quality of predicted keyphrases with respect to the ground truth keyphrases. Our specific evaluation method is based on the symmetric greedy matching embedding-based metric, which was introduced in \cite{rus2012comparison}. 

\section{Approach}

\noindent 
{\bf Recurrent Neural Networks (RNN):} All models investigated in this work are based on RNN, and are inspired by the model proposed by \citeauthor{zhang2016keyphrase} \shortcite{zhang2016keyphrase}. Given an input sequence, $(x_{1},x_{2},x_{3},...,x_{n})$, an RNN processes each token $x_{t}$ at every time-step $t$ by producing a hidden state, $h_t$, using the previous hidden state, $h_{t-1}$, from the previous time-step $t-1$. Formally, this can be expressed as: 
\begin{equation}
h_t = f(x_t,h_{t-1})
\end{equation}
where $f$ is a non-linear activation function.
The final output of an RNN can be a sequence of hidden states, $(h_1, h_2,h_3,...,h_n)$. We used a specific type of RNN, called Long-Short Term Memory (LSTM) network \cite{hochreiter1997long}. An LSTM is equipped with three gates (an input gate, $i$, an output gate, $o$, and a forget gate, $f$), along with a cell-state, $c$, and a hidden state, $h$. The function of an LSTM at each time-step, $t$, given the input, $x_t$, can be described with the following equations: 

\begin{equation}
f_t = \sigma(x_tW_f + h_{t-1}U_f + b_f)
\end{equation}
\begin{equation}
i_t = \sigma(x_tW_i + h_{t-1}U_i + b_i)
\end{equation}
\begin{equation}
o_t = \sigma(x_tW_o + h_{t-1}U_o + b_o)
\end{equation}
\begin{equation}
g_t = \tanh(x_tW_c + h_{t-1}U_c +b_c)
\end{equation}
\begin{equation}
c_t = f_t\odot c_{t-1} + i_t\odot g_t
\end{equation}
\begin{equation}
h_t = o_t\odot \tanh(c_t)
\end{equation}
\setlength{\parskip}{0.1em}

\noindent where $x_t \in \mathbb{R}^{d}, h_t, c_t\in \mathbb{R}^{h}$,  
$W_f, W_i, W_o, W_c$ $\in \mathbb{R}^{d x h}$, 
$U_f, U_i, U_o, U_c$ $\in \mathbb{R}^{h x h}$, 
$b_f,b_i,b_o,b_c \in \mathbb{R}^{h}$
($h$ represents the number of hidden units, and
$d$ represents the dimension of the input), 
$\sigma$ is usually the sigmoid activation function, and $\odot$ is the Hadamard product. 

The LSTM that we used is bidirectional \cite{graves2013hybrid}, in  that it has two encoders: one forward and one backward. The forward encoder starts processing the input sequence $(x_{1},x_{2},x_{3},...,x_{n})$ 
from left to right, thus, creating a forward hidden state representation $(\overrightarrow{h_1}, \overrightarrow{h_2}, \overrightarrow{h_3},..,\overrightarrow{h_n})$ of the input sequence. The backward encoder starts processing the input sequence 
from right to left, thus, creating a backward hidden state representation $(\overleftarrow{h_1}, \overleftarrow{h_2}, \overleftarrow{h_3},..,\overleftarrow{h_n})$ of the input sequence. The final hidden state representation $(h_1, h_2,h_3,...,h_n)$ is created by combining the forward and backward representations (often using the concatenation operator). Formally, we have:
\begin{equation}
\overrightarrow{h_t} = LSTM(x_t,\overrightarrow{h}_{t-1})
\end{equation}
\begin{equation}
\overleftarrow{h_t} = LSTM(x_t,\overleftarrow{h}_{t-1})
\end{equation}
\begin{equation}
h_t = [\overrightarrow{h_t};\overleftarrow{h_t}]
\end{equation}
\noindent To make predictions, linear transformations of the hidden states are then passed to a softmax layer. 

Similar to the model proposed by  \citeauthor{zhang2016keyphrase} \shortcite{zhang2016keyphrase}, our models use a stack of two Bi-LSTMs. The first Bi-LSTM layer takes the original sequence, $(x_{1},x_{2},x_{3},...,x_{n})$, as input, and produces a hidden state sequence, $h_{1:n}^{(1)} = (h_1^{(1)}, h_2^{(1)},h_3^{(1)},...,h_n^{(1)})$. The second Bi-LSTM takes the hidden sequence produced by the first Bi-LSTM layer as  input, and produces a sequence of deeper hidden state representations $h_{1:n}^{(2)} = (h_1^{(2)}, h_2^{(2)},h_3^{(2)},...,h_n^{(2)})$. Formally:
\begin{equation}
h_t^{(1)} = BiLSTM_1(x_t,h_{t-1}^{(1)})
\end{equation}
\begin{equation}
h_t^{(2)} = BiLSTM_2(h_t^{(1)},h_{t-1}^{(2)})
\end{equation}

Following \citeauthor{zhang2016keyphrase} \shortcite{zhang2016keyphrase}, we jointly trained the two stacked Bi-LSTMs on two different but related tasks, in a supervised fashion. The first Bi-LSTM was trained to simply find keywords in a text. This can be viewed as a binary classification task, where detected keywords are classified as `1', and other words are classified as `0'. Let the loss function $J_1(\theta)$ represent the classification error of a non-linear transformation of the first hidden layer units, $h_{1:n}^{(1)}$. The hidden layer of the second Bi-LSTM was trained for a relatively more complex task that builds upon the task of the hidden layer of the first Bi-LSTM. Specifically, the second Bi-LSTM was trained to detect keyphrases by using its hidden units $h_{1:n}^{(2)}$ to tag the sequence using the BIOES scheme. In BIOES, `B' refers to the beginning word of a keyphrase, `E' refers to the last word of the keyphrase, `I' refers to the word being inside the keyphrase (in-between `B' and `E'), `S' refers to any single word keyphrase, and `O' is any word that is not a part of a keyphrase. Let the loss function $J_2(\theta)$ represent the tagging error of a non-linear transformation of the second RNN hidden layer units, $h_{2:n}^{(2)}$. The final loss $J(\theta)$ can then be expressed as the combination of the two losses, $J_1(\theta)$ and $J_2(\theta)$, specifically: 
\begin{equation}
J(\theta) = \gamma\cdot J_1(\theta) + (1-\gamma)\cdot J_2(\theta)
\end{equation}
where $\gamma$ is a hyperparameter in the interval [0,1]. The model updates its trainable parameters to minimize the loss function, $J(\theta)$, through backpropagation during training. 
The loss from a low-level task (keyword discovery) can also serve as additional regularization for the main task (keyphrase extraction).
Intuitively, the two stacked networks  correspond to two different, but related tasks 
- essentially, it is an instance of multi-task learning \cite{liu2016recurrent,sogaard2016deep}. 
\setlength{\parskip}{1em}

\noindent {\bf Contextual Word Embeddings:} Word embeddings are high dimensional vector representations of words. Often, embedding algorithms encode contextual information into the vector representations, such that when clustering word-vectors using the Euclidean distance or Cosine-similarity, vectors corresponding to words that appear in similar contexts tend to cluster together. However, classic word embedding models \cite{pennington2014glove,DBLP:journals/corr/abs-1301-3781} have certain limitations. They can only accommodate one particular vector representation for each word. This restricts the models from being able to capture polysemy (the same word can have completely different meanings in different contexts). The models can also be restricted to a specific vocabulary, which the model has been trained on. Moreover, they can also be blind to morphological information. 
\setlength{\parskip}{0em}

\citeauthor{peters2018deep} \shortcite{peters2018deep} proposed the use of a pre-trained language model to create contextualized word embeddings (called ELMo embeddings), which can tackle some of these issues. They used a character level Convolutional Neural Network (CNN) \cite{zhang2015character} followed by a language model based on Bi-LSTM (however, other architectures can be used \cite{peters2018dissecting}). The word embeddings were constructed by a weighted summation of the hidden state representations of the words at various layers of the language model. As a result, there are some interesting benefits. As the initial word-embedding can be done by a CNN at the character level, there is no word-level vocabulary restriction. Also, as the embedding is done by a language model, the embedding of a word is conditioned on the context of all the surrounding words. Therefore, the same word can be represented using different word-vectors if it appears in different contexts. 

As an enhancement to the main RNN-based model, which uses Glove embeddings\footnote{\url{https://nlp.stanford.edu/projects/glove/}} \cite{pennington2014glove} to represent the input words, 
we concatenated the ELMo embeddings with GloVe embeddings pre-trained on a Twitter corpus. We used Flair\footnote{\url{https://github.com/zalandoresearch/flair}}\cite{akbik2018coling} to load the Twitter GloVe embeddings and TensorFlow hub\footnote{\url{https://tfhub.dev/google/elmo/2}} to load the ELMo embeddings.  \setlength{\parskip}{1em}

\setlength{\textfloatsep}{1em}
\begin{figure}
    \centering
\tikzset{every picture/.style={line width=0.75pt}} 

\begin{tikzpicture}[x=0.75pt,y=0.75pt,yscale=-0.75,xscale=0.9]

\draw   (38,581.68) -- (113.3,581.68) -- (113.3,615.5) -- (38,615.5) -- cycle ;
\draw   (134,581.37) -- (209.3,581.37) -- (209.3,615.19) -- (134,615.19) -- cycle ;
\draw   (228,581.2) -- (303.3,581.2) -- (303.3,615.02) -- (228,615.02) -- cycle ;
\draw   (37.5,524.4) -- (112.8,524.4) -- (112.8,562.02) -- (37.5,562.02) -- cycle ;
\draw    (76.3,581.2) -- (76.17,564.6) ;
\draw [shift={(76.15,562.6)}, rotate = 449.54] [color={rgb, 255:red, 0; green, 0; blue, 0 }  ][line width=0.75]    (10.93,-3.29) .. controls (6.95,-1.4) and (3.31,-0.3) .. (0,0) .. controls (3.31,0.3) and (6.95,1.4) .. (10.93,3.29)   ;

\draw   (37.5,471.7) -- (112.8,471.7) -- (112.8,505.52) -- (37.5,505.52) -- cycle ;
\draw   (148,489.77) .. controls (148,477.12) and (158.25,466.87) .. (170.9,466.87) .. controls (183.55,466.87) and (193.8,477.12) .. (193.8,489.77) .. controls (193.8,502.41) and (183.55,512.67) .. (170.9,512.67) .. controls (158.25,512.67) and (148,502.41) .. (148,489.77) -- cycle ;
\draw   (133,410.2) -- (208.3,410.2) -- (208.3,444.02) -- (133,444.02) -- cycle ;
\draw   (134.7,346.58) -- (210,346.58) -- (210,380.4) -- (134.7,380.4) -- cycle ;
\draw   (35,376.8) -- (110.3,376.8) -- (110.3,416.8) -- (35,416.8) -- cycle ;
\draw   (133,226.2) -- (208.3,226.2) -- (208.3,266.4) -- (133,266.4) -- cycle ;
\draw    (170.3,346.4) -- (170.3,328.2) ;
\draw [shift={(170.3,326.2)}, rotate = 450] [color={rgb, 255:red, 0; green, 0; blue, 0 }  ][line width=0.75]    (10.93,-3.29) .. controls (6.95,-1.4) and (3.31,-0.3) .. (0,0) .. controls (3.31,0.3) and (6.95,1.4) .. (10.93,3.29)   ;

\draw    (170.3,393.7) -- (115.3,394.18) ;
\draw [shift={(113.3,394.2)}, rotate = 359.5] [color={rgb, 255:red, 0; green, 0; blue, 0 }  ][line width=0.75]    (10.93,-3.29) .. controls (6.95,-1.4) and (3.31,-0.3) .. (0,0) .. controls (3.31,0.3) and (6.95,1.4) .. (10.93,3.29)   ;

\draw   (133,286.8) -- (208.3,286.8) -- (208.3,326.8) -- (133,326.8) -- cycle ;
\draw   (33,313.2) -- (108.3,313.2) -- (108.3,353.4) -- (33,353.4) -- cycle ;
\draw    (169.9,468.2) -- (170.27,444.2) ;
\draw [shift={(170.3,442.2)}, rotate = 450.88] [color={rgb, 255:red, 0; green, 0; blue, 0 }  ][line width=0.75]    (10.93,-3.29) .. controls (6.95,-1.4) and (3.31,-0.3) .. (0,0) .. controls (3.31,0.3) and (6.95,1.4) .. (10.93,3.29)   ;

\draw    (71.3,375.2) -- (71.3,355.2) ;
\draw [shift={(71.3,353.2)}, rotate = 450] [color={rgb, 255:red, 0; green, 0; blue, 0 }  ][line width=0.75]    (10.93,-3.29) .. controls (6.95,-1.4) and (3.31,-0.3) .. (0,0) .. controls (3.31,0.3) and (6.95,1.4) .. (10.93,3.29)   ;

\draw    (170.3,408.2) -- (170.3,381.2) ;
\draw [shift={(170.3,379.2)}, rotate = 450] [color={rgb, 255:red, 0; green, 0; blue, 0 }  ][line width=0.75]    (10.93,-3.29) .. controls (6.95,-1.4) and (3.31,-0.3) .. (0,0) .. controls (3.31,0.3) and (6.95,1.4) .. (10.93,3.29)   ;

\draw    (170.3,286.4) -- (170.3,268.2) ;
\draw [shift={(170.3,266.2)}, rotate = 450] [color={rgb, 255:red, 0; green, 0; blue, 0 }  ][line width=0.75]    (10.93,-3.29) .. controls (6.95,-1.4) and (3.31,-0.3) .. (0,0) .. controls (3.31,0.3) and (6.95,1.4) .. (10.93,3.29)   ;

\draw    (169.65,657) -- (169.32,616.6) ;
\draw [shift={(169.3,614.6)}, rotate = 449.53] [color={rgb, 255:red, 0; green, 0; blue, 0 }  ][line width=0.75]    (10.93,-3.29) .. controls (6.95,-1.4) and (3.31,-0.3) .. (0,0) .. controls (3.31,0.3) and (6.95,1.4) .. (10.93,3.29)   ;

\draw    (169.65,657) -- (80.97,616.24) ;
\draw [shift={(79.15,615.4)}, rotate = 384.69] [color={rgb, 255:red, 0; green, 0; blue, 0 }  ][line width=0.75]    (10.93,-3.29) .. controls (6.95,-1.4) and (3.31,-0.3) .. (0,0) .. controls (3.31,0.3) and (6.95,1.4) .. (10.93,3.29)   ;

\draw    (169.65,657) -- (264.32,615.7) ;
\draw [shift={(266.15,614.9)}, rotate = 516.4300000000001] [color={rgb, 255:red, 0; green, 0; blue, 0 }  ][line width=0.75]    (10.93,-3.29) .. controls (6.95,-1.4) and (3.31,-0.3) .. (0,0) .. controls (3.31,0.3) and (6.95,1.4) .. (10.93,3.29)   ;

\draw    (75.8,524.2) -- (75.67,507.6) ;
\draw [shift={(75.65,505.6)}, rotate = 449.54] [color={rgb, 255:red, 0; green, 0; blue, 0 }  ][line width=0.75]    (10.93,-3.29) .. controls (6.95,-1.4) and (3.31,-0.3) .. (0,0) .. controls (3.31,0.3) and (6.95,1.4) .. (10.93,3.29)   ;

\draw    (170.87,579.73) -- (170.9,514.67) ;
\draw [shift={(170.9,512.67)}, rotate = 450.03] [color={rgb, 255:red, 0; green, 0; blue, 0 }  ][line width=0.75]    (10.93,-3.29) .. controls (6.95,-1.4) and (3.31,-0.3) .. (0,0) .. controls (3.31,0.3) and (6.95,1.4) .. (10.93,3.29)   ;

\draw    (112.87,490.4) -- (146,489.8) ;
\draw [shift={(148,489.77)}, rotate = 538.97] [color={rgb, 255:red, 0; green, 0; blue, 0 }  ][line width=0.75]    (10.93,-3.29) .. controls (6.95,-1.4) and (3.31,-0.3) .. (0,0) .. controls (3.31,0.3) and (6.95,1.4) .. (10.93,3.29)   ;

\draw    (271.53,582.4) -- (272.84,491.73) ;
\draw [shift={(272.87,489.73)}, rotate = 450.82] [color={rgb, 255:red, 0; green, 0; blue, 0 }  ][line width=0.75]    (10.93,-3.29) .. controls (6.95,-1.4) and (3.31,-0.3) .. (0,0) .. controls (3.31,0.3) and (6.95,1.4) .. (10.93,3.29)   ;

\draw    (272.87,489.73) -- (195.8,489.77) ;
\draw [shift={(193.8,489.77)}, rotate = 359.98] [color={rgb, 255:red, 0; green, 0; blue, 0 }  ][line width=0.75]    (10.93,-3.29) .. controls (6.95,-1.4) and (3.31,-0.3) .. (0,0) .. controls (3.31,0.3) and (6.95,1.4) .. (10.93,3.29)   ;

\draw (169.5,663.5) node  [align=left] {{\footnotesize `harvey'}};
\draw (75.5,598) node [scale=0.8,xslant=0.04] [align=left] {IPA\&Phonology\\Embeddings};
\draw (171,597) node [scale=0.8,xslant=0.04] [align=left] {Word \\Embeddings};
\draw (76,543) node  [align=left] {{\footnotesize Char-level }\\{\footnotesize CNN}};
\draw (267,599) node [scale=0.8,xslant=0.04] [align=left] {POS \\Embeddings};
\draw (76,489.5) node  [align=left] {{\footnotesize CNN output}};
\draw (171.67,490) node  [align=left] {{\footnotesize Concat}};
\draw (171,426) node  [align=left] {{\footnotesize Bi-LSTM 1}};
\draw (172,363) node  [align=left] {{\footnotesize Bi-LSTM 2}};
\draw (171,307) node  [align=left] {{\footnotesize Linear + }\\{\footnotesize Softmax}};
\draw (72,395) node  [align=left] {{\footnotesize Linear + }\\{\footnotesize Softmax}};
\draw (171,247) node  [align=left] {{\footnotesize Keyphrase }\\{\footnotesize Extraction}};
\draw (70,334) node  [align=left] {{\footnotesize Keyword }\\{\footnotesize Discovery}};
\draw (210.5,663.5) node  [align=left] {{\footnotesize `hurricane'}};
\draw (239.5,663.5) node  [align=left] {{\footnotesize `is'}};
\end{tikzpicture}
\caption{Joint-layer Bi-LSTM with word embeddings, POS-tags embeddings, and IPA \& phonological embeddings. 
} \label{fig:1}
\end{figure}
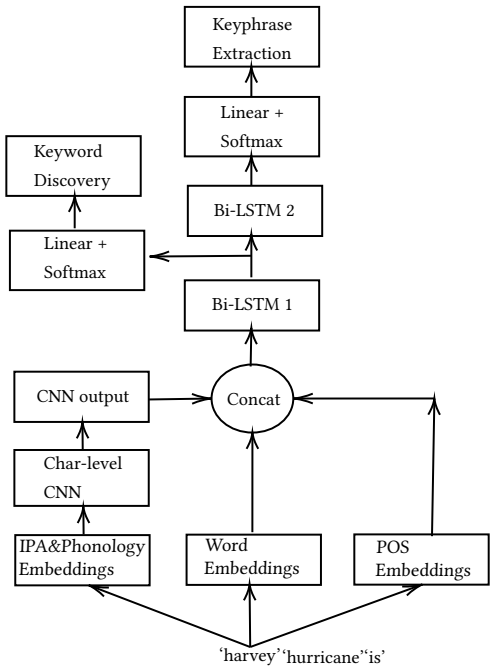

\noindent \textbf{Modeling Tweet Noisiness:} In Twitter, people often use shorthands, slang, arbitrary capitalizations, and varied lexical representations for a word, all of which contribute to a high overall noisiness in tweets. For this reason, many Natural Language Processing tools are found to perform noticeably worse when used on Twitter or similar social media domains, even when they perform very well on cleaner data (like news articles) \cite{ritter2011named, aguilar2018modeling}. \citeauthor{aguilar2018modeling} \cite{aguilar2018modeling} proposed various ways to combat the inherent noisiness in social media platforms. They noted how Twitter users often tend to spell words based on their pronunciations. Thus, while spellings of a word may differ in their lexical format, they tend to have similar phonetics. Therefore, we can create more normalized representations of the words by using their phonetics or corresponding IPA (International Phonetic Alphabet) letters, alongside with their phonological features. 
Following the work of \citeauthor{aguilar2018modeling} \shortcite{aguilar2018modeling}, we used Epitran\footnote{\url{https://github.com/dmort27/epitran}} \cite{Mortensen-et-al:2018} to convert graphemes to phonemes (represented using IPA), and Panphon\footnote{\url{https://github.com/dmort27/panphon}} \cite{Mortensen-et-al:2016} to convert each IPA phoneme into a vector representing various phonological (articulatory) features associated with it. In addition, we noticed that words that are nouns or adjectives are generally better candidates for keywords. For example, disaster names (nouns) can often be used as keyphrases. Therefore, similar to \cite{aguilar2018modeling}, we also added part-of-speech (POS) information to our model. We used the Twitter optimized POS-tagger\footnote{\url{http://www.cs.cmu.edu/~ark/TweetNLP/}} as provided by \citeauthor{owoputi2013improved} \cite{owoputi2013improved} for POS-tagging. \setlength{\parskip}{0em}

We used randomly initialized trainable embeddings for each POS-tag and IPA symbol. We directly used the phonological vector representations created with Panphon as embeddings for phonological features. We concatenated the embeddings of IPA symbols with their corresponding phonological feature vectors. The result of the concatenation, a character level representation of the phonetics and phonological features for each word, was fed to a character-level CNN \cite{zhang2015character}, followed by a global max-pooling layer, to create word level representations. 
Note that the character-level processing may also allow the network to capture some morphological information. Unlike \citeauthor{aguilar2018modeling}, we chose to use a CNN as opposed to a Bi-LSTM for creating the word level representations, because using a Bi-LSTM at character level can be much more computationally expensive and time-consuming than using a CNN. The output of the CNN was concatenated with the POS embeddings, ELMo embeddings, and GloVe embeddings. The resultant representation was then fed to the stacked Bi-LSTM. Figure \ref{fig:1} illustrates this setup.
\setlength{\parskip}{0em}

\section{Datasets}

\noindent We performed our experiments on both general and disaster-related Twitter datasets. For the general Twitter dataset, we used the  dataset\footnote{\url{https://github.com/fudannlp16/KeyPhrase-Extraction}} published in \cite{zhang2016keyphrase}, which is annotated with keyprases. The training and test data were provided as separate files. We used $80\%$ of the training data for actual training and 20\%  for validation. 

We constructed the disaster training and validation datasets by combining disaster-related tweets that have hashtags, from several disasters, specifically, Boston Bombing,  Hurricane Harvey, and  Hurricane Sandy. We used $80\%$ of the data for training and 20\%  for validation. We used the combination of disaster-related tweets, with hashtags, from Mexico Earthquake,  Hurricane Maria, and Hurricane Irma as test data. To annotate the training, validation and test tweets with keyphrases, we first made use of hashtags. The hashtagged units (marked with `\#') in the tweets used are often multi-word phrases without any spaces in-between. We used WordSegment\footnote{\url{http://www.grantjenks.com/docs/wordsegment/}} to split the hashtagged multi-word phrases into individual words. These segmented words were used as ground truth for keyphrase extraction. This is similar to how \citeauthor{zhang2016keyphrase} \cite{zhang2016keyphrase} prepared their annotations. Since hashtags are often used to mark topical phrases or phrases related to some trends, they represent good candidates for keyphrases.  In addition, we also used a crisis-lexicon \cite{olteanu2014crisislex} to find disaster-related words (at both unigram and bigram level) in the dataset, and marked them as ground-truth keyphrases. We also added some of our own phrases such as ``Boston Bombing'', ``Hurricane Sandy'', ``Mexico Earthquake'' to the crisis-lexicon. 

For both the general and disaster datasets, we filtered out tweets that were longer than 200 tokens or shorter than 5 tokens after tokenization. Table \ref{table: 3} shows the exact number of tweets we used for training, validation, and testing in each dataset. In addition,  Table \ref{table: 3} shows the total number of keyphrases and the average number of keyphrases per tweet in each dataset. As can be seen, the average number of keyphrases/hashtags per tweet in the disaster dataset is approximately twice the average number of keyphrases/hashtags per tweet. 

\setlength{\textfloatsep}{0em}
\begin{table}[t]
\small
\centering
\def\arraystretch{1.2}
\begin{tabular}{ | p{7em} | p{12em} | p{3em} | } 
\hline
General Twitter \newline\newline & 
Training Samples \newline
Validation Samples \newline
Test Samples: \newline
Total Keyphrases \newline
Avg Keyphrases per Tweet \newline
& 62290 \newline 
15572 \newline
33392 \newline
111254 \newline
1.00  \\
\hline
Disaster Twitter \newline\newline & 
Training Samples \newline
Validation Samples \newline
Test Samples\newline
Total Keyphrases \newline
Avg Keyphrases per Tweet \newline
& 68128 \newline 
17032 \newline
37106 \newline
221408 \newline 
1.81\\
\hline
\end{tabular}
\vspace{+.5em}
\caption{ Dataset summary.}
\label{table: 3}
\end{table}

\begin{figure*}[t]
  \centering
 \subfloat[General test data] 
 {\includegraphics[width=28em]{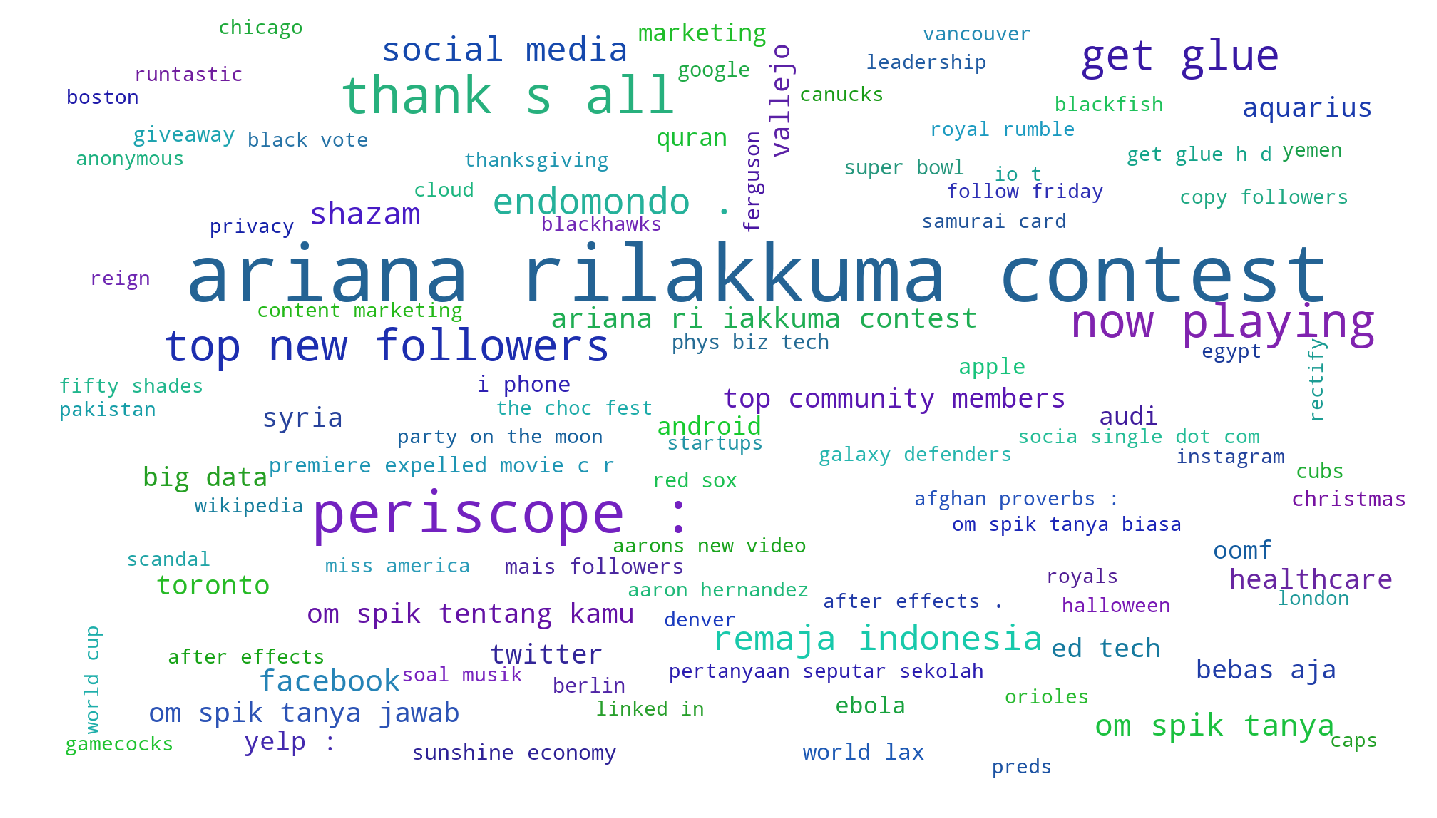}}
   \subfloat[Disaster test data]{\includegraphics[width=28em]{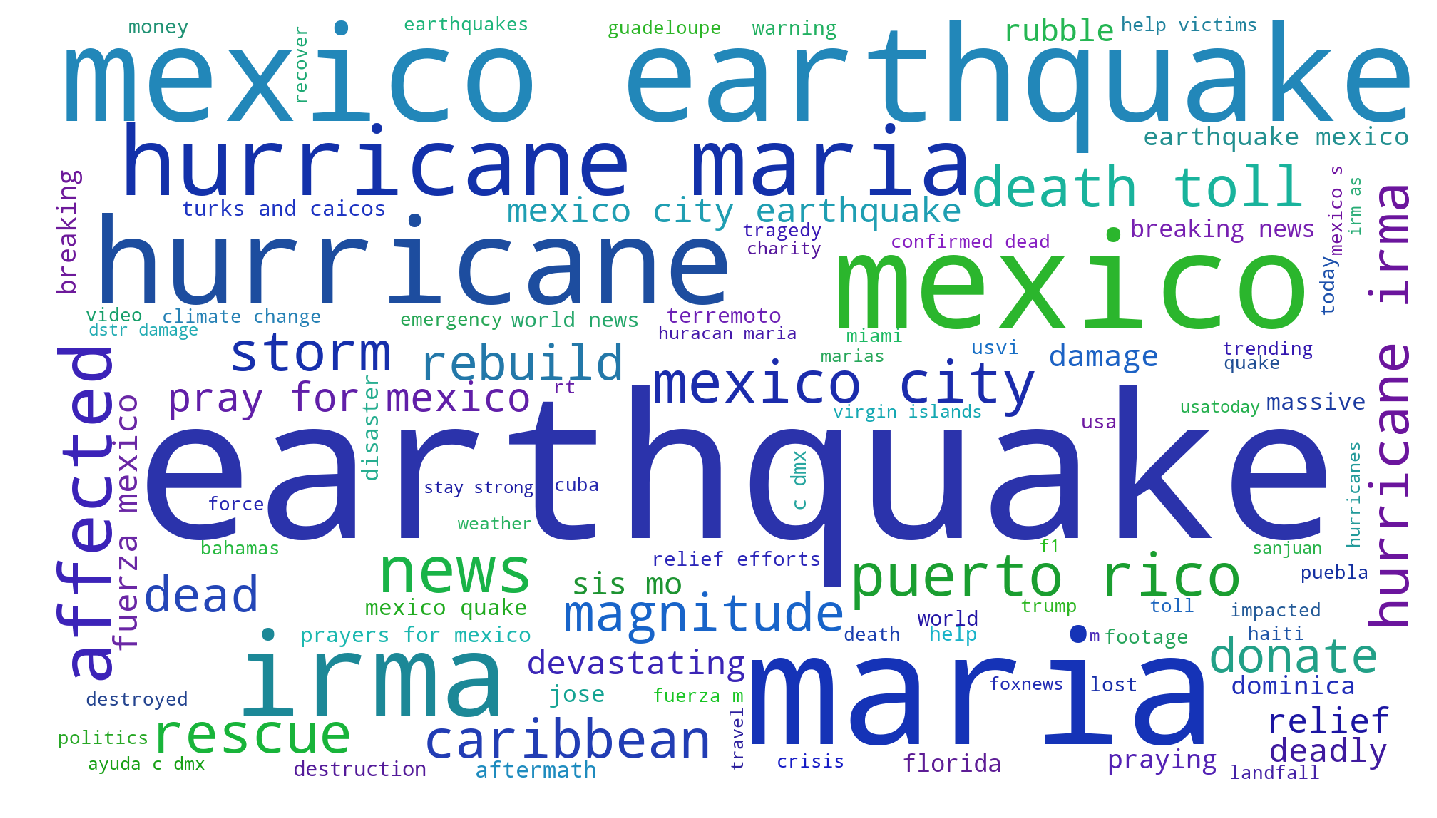}}
  \caption{Word clouds of the top 100 most frequent gold keyphrases from the
   general test data (a) and disaster test data (b).}
  \label{fig:2}
\end{figure*}

There are other differences in terms of keyphrase usage between the general and the disaster datasets.  First, the types of keyphrases used in the two datasets are quite different, as the disaster dataset contains a more limited vocabulary related to disasters, while the general dataset has a more diverse vocabulary. This can be seen by comparing the two wordclouds shown in Figure \ref{fig:2}. Furthermore, from Figure \ref{fig:2}, we can also see that there are many frequently occurring keyphrases in the disaster data, whereas there are a lot fewer keyphrases that appear as frequently in the general data. At last, we point out that the disaster test data is quite different from the disaster training (and validation) data, given that the test data contains tweets related to different disasters as compared to the disasters in the training and validation data. In particular, a significant fraction of the disaster test data is made of disaster tweets posted during the Mexico Earthquake, while the training and validation data do not contain tweets related to earthquakes. However, both training/validation and test data contain tweets collected during hurricanes, although the hurricanes used in the training/validation are different from those used in the test. The differences in the disaster training and test data can be observed in Figure \ref{fig:3}. We intentionally created a challenging test dataset that allows us to evaluate the model's performance on data from a disaster that does not appear in the training, to understand the usability of the model in a realistic scenario.
\setlength{\parskip}{0em}

\section{Experiments and Results}

\setlength{\parskip}{0.5em}
\subsection{Experimental Setting} 
\label{setting}
\setlength{\parskip}{0.3em}

One of our goals was to enhance the original model proposed in \cite{zhang2016keyphrase} with contextual information (in the form of contextual word embeddings), and also with information that can alleviate the noisiness of the Twitter data (specifically, phonology and POS tag information). Thus, our set of experiments was designed to compare the baseline model \cite{zhang2016keyphrase} with the proposed variants.  The implementation details of the models  are described below. Hyper-parameters were selected based on common values used in similar prior architectures, and/or (limited) tuning on the validation set. \setlength{\parskip}{1em}

\noindent
{\bf Joint-layer Bi-LSTM} (J-Bi-LSTM): This model serves as our baseline. We parsed each sequence using a window of size 3, that is, each token in the sequence was represented by three consecutive words. As an example, if the sequence is $(s_1,s_2,s_3)$, after parsing it with a window of size 3, the sequence gets converted into $([SOS,s_1,s_2], [s_1,s_2,s_3], [s_2,s_3,EOS])$ (where SOS is a special tag denoting the start of the sequence, and EOS is a similar tag denoting the end of the sequence). The 3-word representation of a token was used to encode more contextual information into each token, before feeding the tokens into the LSTM. \citeauthor{zhang2016keyphrase} \cite{zhang2016keyphrase} empirically found 3 to be a good choice for the window size. Each word/token in the sequence was embedded with a 100-dimensional GloVe vector pre-trained on a general Twitter corpus. The embeddings were further fine-tuned during training. Each LSTM network had 300 hidden units. We used a dropout of 0.5 \cite{baldi2013understanding} on the input to the first and second LSTMs. We also used a dropout with the same rate on the output from the second LSTM. For training, we used the nadam optimizer \cite{dozat2016incorporating} with a learning rate of 0.0015. Finally, we used an $\gamma$ value of 0.5 for joining the two loss functions $J_1(\theta)$ and $J_2(\theta)$. Tuning this hyperparameter may lead to better results.


\noindent
{\bf J-Bi-LSTM with ELMo Embeddings}  (J-Bi-LSTM+ELMo): This variant is similar to the baseline model, except that it uses 1024 dimensional ELMo embeddings, along with the GloVe embeddings. The two types of embeddings, GloVe and ELMo, are concatenated together. As the ELMo embeddings have much higher dimensionality, and are designed to  encode contextual information, we did not use the window approach for this model. Other implementation details and hyperparameters are similar to those in the baseline. 

\noindent
{\bf J-Bi-LSTM with IPA \& POS} (J-Bi-LSTM+IPA+POS): This variant is also similar to the baseline, except that it uses IPA-embeddings, phonological
feature vectors and POS-tag embeddings along with the GloVe embeddings (with a window of size 3). We used 64-dimensional embeddings for POS-tags and 22-dimensional embeddings for IPA symbols. The phonological feature vectors have 22 dimensions as well. For the character-level CNN, we used 128 filters and a kernel of size 3. As before, other implementation details and hyperparameters are similar to those in the baseline. 

\noindent
{\bf J-Bi-LSTM with ELMo, IPA \& POS} (J-Bi-LSTM+ELMo+IPA+POS): To evaluate the combined effect of the contextual word embeddings and IPA/POS-tags, this variant makes use of IPA-embeddings, phonological feature vectors and POS-tag embeddings, along with the concatenation of GloVe and ELMo embeddings. Other implementation details and hyperparameters are the same as before. 



\noindent We compare the above models in three different settings: (1) we train all models on general tweet data and test them also on general tweet data. The goal of this comparison was to see if the additional information used can improve the performance of the baseline; (2) we train all models on the general tweet data and test them on the disaster tweet data, with the goal of understanding how much information can be transferred from the general tweet data to the disaster-specific tweet data; (3) we train all models on disaster tweet data and test them also on disaster tweet data. The goal of this experiment was to see how much the models improve by being trained on specific disaster data, and also what type of information leads to the largest performance improvements.\setlength{\parskip}{0em}





\subsection{Evaluation Metrics}
\label{metrics}
\setlength{\parskip}{0.3em}

The J-Bi-LSTM based models take a sequence-labeling approach towards keyphrase extraction. Thus, the precision, recall, and F1-measure can be used to evaluate the results. 
A keyphrase predicted by the model is considered to be correct if there exists an exact matching keyphrase in the same position in the ground-truth annotation; otherwise, the prediction is considered incorrect. 



\setlength{\parskip}{0.3em}

\noindent Despite being frequently used, the F1-measure may not be ideal for evaluating the quality of predicted keyphrases. For example, if the ground-truth keyphrase is ``harvey'', and the extracted keyphrase is ``harvey hurricane'', the exact-F1 measure will be zero. Therefore, the model can produce appropriate predictions, or sometimes even better predictions than the ground-truth annotations, but still, receives a zero score. Similarly, a model may detect some keyphrases that are semantically similar to the ground-truth, but not exact matches. For example, the model may detect `tremor' instead of `earthquake'. In these cases too, the model will receive a zero score, whereas presumably some points should be awarded to the model for predicting something close to the ground truth. \setlength{\parskip}{0em}

The above reasons led us to explore embedding-based measures. It has been shown that the cosine-similarity between certain vector representations (such as GloVe or word2vec embeddings) of semantically similar words tend to be high. Intuitively, we can use cosine-similarity of such word embeddings to measure the degree of similarity between them, as an alternative to simply checking for an exact match. 
Embedding-based metrics have been used in evaluating the quality of generated dialogues, detecting paraphrases, assessing student responses, and other similar tasks. Here, we are interested in comparing the ground-truth keyphrases with the predicted keyphrases. However, taking the order and positional information into account, while comparing them can be complicated. For example, consider a case where there are two gold keyphrases `$K_1$' and `$K_2$' at consecutive locations $P_1$ and $P_2$, respectively, whereas the predicted keyphrase is `$K_1$ $K_2$' (covering both $P_1$ and $P_2$ locations). Such cases make it complicated to use positional information for comparison. Thus, we moved our focus from the standard sequence-labeling evaluation approach to an approach that seems more natural to the task of keyphrase extraction. 

Specifically, we first extracted the gold and  predicted keyphrases from a labeled sequence and converted them into sets of keyphrases. By doing this, the order information was ignored, and duplicates were removed. As a result, we obtained two sets - a set of gold keyphrases and a set of predicted keyphrases. 
Our goal was to use the embedding based metrics to measure the overlap between these sets. However, each keyphrase in the gold and predicted sets, respectively, can have an arbitrary number of words, each with its own word embedding. This further complicates the keyphrase-to-keyphrase comparison. To simplify the task, we took the average of the word-embeddings in each phrase to create phrase-level embeddings. Given the phrase-level embeddings, the embedding-based metrics can be used in several ways to score the overall quality of the predicted keyphrases using the gold keyphrases as the standard. Some of these methods are described below. 
\setlength{\parskip}{1em}

\noindent \textbf{Average Embeddings:} One approach is to take the average \cite{foltz1998measurement, landauer1997solution, mitchell2008vector} of all phrase-level embeddings in the set of gold keyphrases. (Alternatively, we can also use the most extreme value \cite{forgues2014bootstrapping} along each dimension, across all the phrase-level-embeddings in the set.) Then, we can either compute the average of cosine-similarity scores between the phrase-level embedding of each predicted keyphrase and the average-gold-embedding, or we can compute the cosine-similarity score between the average embedding of the predicted keyphrases and the average embedding of the gold keyphrases. 


\noindent \textbf{Greedy Matching:} Another approach is to make word to word (or in our case, keyphrase to keyphrase) comparisons without creating single representations for all keyphrases in the sets. In this approach, we can score each keyphrase from the set of predicted keyphrases in two steps. In the first step, we compute the cosine-similarity scores between the predicted keyphrase under consideration and each keyphrase in the set of gold keyphrases. In the second step, we take the maximum of all the computed cosine-similarity scores for each predicted keyphrase. The maximum cosine-similarity score should correspond to the best matching keyphrase in the gold set for a particular predicted keyphrase. Thus, each predicted keyphrase can be scored. Finally, we can take the average of the scores to get the final score for the model's prediction \cite{rus2012comparison}: \setlength{\parskip}{0em} 

\begin{equation}
GM(\, pred,gold \,) = \frac{\sum_{i=1}^p score(pred_i, gold)}{p}   
\end{equation}

\setlength{\parskip}{0.3em}
\noindent
where $p$ is the number of predicted keyphrases, {\it pred} is the set of predicted keyphrases, {\it gold} is the set of gold keyphrases, and $pred_i$, with $i \in \overline{1,p}$, are the elements of {\it pred}. {\it GM} is the greedy match scoring function, and $score(pred_i, gold)$ is the maximum cosine-similarity achieved between $pred_i$ and any keyphrase from {\it gold}. \setlength{\parskip}{0em}

However, this scoring is not symmetric. $GM(\, pred,gold \,)$ will not necessarily be the same as $GM(\, gold,pred \,)$. Inspired from  
\cite{rus2012comparison}, we can use the following formula to calculate a symmetric score: \setlength{\parskip}{0em}

\begin{equation}
T = GM(\, pred,gold \,) + GM(\, gold,pred \,)
\end{equation}
\begin{equation}
SymmGM(\, pred,gold \,) = \frac{T}{2}  
\end{equation}

\setlength{\parskip}{0.3em}

\noindent There is another benefit of the symmetric scoring. If $GM( pred,gold)$ is used alone, the model can be scored highly even if there is only one prediction in the set which happens to match very well with just one gold keyphrase, among multiple other gold keyphrases that were not predicted. Similarly, $GM(\, gold,pred \,)$ can produce a high score, even if there is only one keyprhase in the gold set that matches very well with just one predicted keyphrase, among multiple other predicted keyphrases that do not match with the gold. Using  $SymmGM(\, pred,gold \,)$ mitigates these issues. \setlength{\parskip}{1em}

\noindent \textbf{Optimal matching:} This approach, proposed by \citeauthor{rus2012comparison} \cite{rus2012comparison}, can be used to first create an optimal group of exclusive phrase-to-phrase pairs using the Kuhn-Munkres method \cite{kuhn} for optimal assignment. Then, each predicted keyphrase can be scored simply by computing its cosine-similarity with the gold-keyphrase that it is paired with. The later steps are similar to those of greedy matching. 

\noindent \textbf{Proposed embedding-based metric:}  We chose greedy matching as the backbone of our proposed embedding-based metric. As we are dealing with small sets of discrete keyphrases, we found it more appropriate to choose a phrase-to-phrase comparison approach such as greedy matching or optimal matching. A predicted keyphrase can be considered appropriate if it is similar to any of the gold keyphrases. Therefore, to score the correctness of the predictions, we chose to allow multiple predicted keyphrases to be matched to the same gold keyphrase if needed. However, optimal matching can only allow exclusive one-to-one pairs. This is why we did not choose the optimal matching. \setlength{\parskip}{0em}

\begin{table*}[t]
\small
\centering
\def\arraystretch{1.2}
\begin{tabular}{| l | p{29em} | c | p{7em} | p{7em} | p{3.3em}| p{3em}|}
\hline
& \textbf{Gold and prediction} & \textbf{F1} & \textbf{No alpha-beta \newline No threshold \newline General Embd} & \textbf{No alpha-beta \newline No threshold \newline Crisis Embd} & \textbf{General \newline Embd} & \textbf{Crisis\newline Embd} \\
\hline
1 &
\textbf{Gold:} teen choice\newline
\textbf{Prediction:} teen choice awards & 0\% & 95.6\% & 88.4\%  & 95.6\% & 88.4\%\\
\hline
2 &
\textbf{Gold:} earthquake, volunteers, working together \newline
\textbf{Prediction:} earthquake, volunteers working together
& 40\% & 93.1\% & 92.1\% & 80.5\% & 79.8\%\\
\hline
3 &
\textbf{Gold:} storm, cloud, calamity \newline
\textbf{Prediction:} tornado, cloudy sky, disaster
& 0\% & 61.1\% & 30.9\% & 52.3\% & 13.4\%\\
\hline
4 &
\textbf{Gold:} harvey \newline
\textbf{Prediction:} hurricane
& 0\% & 25.2\% & 63.20\% & 0\% & 63.20\%\\
\hline
5 & 
\textbf{Gold:} hurricane harvey, hurricane, katrina, red cross, bombing \newline
\textbf{Prediction:} hurricane, harvey, katrina hurricane, cross, boston bombing
& 20\% & 90\% & 93.4\% & 90\% & 93.4\%\\
\hline
6 & 
\textbf{Gold:} boston bombing, hurricane harvey, red cross \newline
\textbf{Prediction:} hurricane harvey
& 50\% & 83.3\% & 77.4\% & 54.2\% & 44.4\%\\
\hline
7 & 
\textbf{Gold:} boston bombing, hurricane harvey \newline
\textbf{Prediction:} hurricane harvey, boston, bombing, red cross
& 33.3\% & 90.2\% & 90.3\% & 70.3\% & 70.2\%\\
\hline
\end{tabular}
\vspace{0.5em}
\caption{General and crisis embedding-based scores versus F1 scores on seven sets of gold/predicted keyphrases.}
\vspace{-5mm}
\label{table: 4}
\end{table*}

We should note that if we simply use the average score when calculating $GM(\, pred,gold \,)$ and $GM(\, gold,pred \,)$, the variation in the number of keyphrases in {\it gold} and {\it pred} may not be completely taken into account by greedy matching. More specifically, $GM(\, pred,gold \,)$ can be high as long as all the predicted keyphrases achieve high cosine-similarity scores with any one of the gold keyphrases. But it may not take into account how many more or less keyphrases there may be in the {\it gold} set as compared to those in the {\it pred} set. Similarly, $GM(\, gold,pred \,)$ can be high, as long as all the gold keyphrases achieve high cosine-similarity scores with any of the predicted keyphrases, without taking into account how many more or less keyphrases there may be in the {\it pred} set as compared to those in the {\it gold} set. Both $GM(\, pred,gold \,)$ and $GM(\, gold,pred \,)$ may be high at the same time despite wide variation in the number of predicted keyphrases and gold keyphrases, 
when most of the keyphrases in {\it gold} and {\it pred} are very similar to each other. 
To account for this, we may want to penalize the model for extracting too little or too many keyphrases compared to the gold. This can be done with a simple modification to the $GM$ definition:

\vspace{-3mm}
\begin{equation}
GM'(\, pred,gold \,) = \frac{\sum_{i=1}^p score(pred_i, gold)}{max(p,g)}  
\end{equation}

\begin{equation}
GM'(\, gold,pred \,) = \frac{\sum_{i=1}^g score(gold_i, pred)}{max(p,g)}
\end{equation}

\begin{equation}
T' = GM'(\, pred,gold \,) + GM'(\, gold,pred \,)
\end{equation}
\vspace{-3mm}
\begin{equation}
SymmGM'(\, pred,gold \,) = \frac{T'}{2} 
\end{equation}
\setlength{\parskip}{0.3em}
\vspace{-3mm}

\noindent where $g$ is the number of gold keyphrases, and $p$ is the number of predicted keyphrases as defined before. $SymmGM'(\, pred,gold \,)$ is the overall score. If $g$ is higher than $p$, the $GM'(pred,gold)$ score will be lower as compared to the $GM(pred,gold)$ score, and if $p$ is higher than $g$, then $GM'(\, gold,pred \,)$ will be lower as compared to $GM(\,gold,pred \,)$. 
Intuitively, we can think of $GM'(pred,gold)$ as penalizing the overall score, when $g$ is higher than $p$, and we can think of $GM'(\, gold,pred \,)$ as penalizing the overall score when $p$ is higher than $g$. 
\setlength{\parskip}{0em}

However, we may also want to have more fine-grained control over the degree of penalty. Typically, $SymmGM(\, pred,gold \,)$ will be high even when there is wide variation in the lengths of {\it pred} and {\it gold} sets if most of the predicted and gold kephrases are similar to each other. But if that is the case, the predictions are still likely to be appropriate. Moreover, in general, variation in the number of keyphrases can be implicitly penalized by plain averages (with $GM(\,gold,pred \,)$ and  $GM(\,pred,gold \,)$) because with wide variation, there will be higher chances for there to be phrases unique to only one set, which are dissimilar with any of the phrases in the other set. 
For these reasons, we may want to constrain the penalty for the variation of length, which can be achieved as another extension of the $GM$ scoring. 
First note that: \setlength{\parskip}{0em}
\begin{equation}
max(p,g) = p + max(0,g-p) = g + max(0,p-g)
\end{equation}
\noindent
To gain more fine-grained control over the penalty for the length variation, we use two new parameters $\alpha$ and $\beta$, as follows:

\begin{equation}
GM_{ext}(\, pred,gold \,) = \frac{\sum_{i=1}^p score(pred_i, gold)}{p + \alpha\cdot max(0,g-p)}
\end{equation}

\begin{equation}
GM_{ext}(\, gold,pred \,) = \frac{\sum_{i=1}^g score(gold_i, pred)}{g + \beta\cdot max(0,p-g)}
\end{equation}

\begin{equation}
T_{ext} = GM_{ext}(\, pred,gold \,) + GM_{ext}(\, gold,pred \,)
\end{equation}

\begin{equation}
G(\, pred,gold \,) = \frac{T_{ext}}{2}
\end{equation}
\setlength{\parskip}{0.3em}
\vspace{-3mm}

\noindent Intuitively, $\alpha$ and $\beta$ can be used to modulate the penalty for $p$ being smaller than $g$, and for $p$ being greater than $g$, respectively. This leads to a more general formulation. For example, when $\alpha$ is 0, $GM_{ext}(\, pred,gold \,)$ is just $GM(\, pred,gold \,)$, and when $\alpha$ is 1, $GM_{ext}(\, pred,gold \,)$ is just $GM'(\, pred,gold \,)$. $G(\, pred,gold \,)$ is symmetric when $\alpha$ and $\beta$ have the same value. \setlength{\parskip}{0em}

We used $G(\, pred,gold \,)$ to compare a predicted keyphrase set with a gold keyphrase set, with $\alpha$ and $\beta$ both being set to 0.7. In addition, we also used a threshold $\theta$ of 0.4 on each cosine-similarity scores to limit the contribution of lower-scoring matches, and implicitly, to limit the range of $G(\, pred,gold \,)$ to values between 0 and 1. 
That is, if the cosine similarity score 
is lower than a specified threshold, we assigned it a score of 0, otherwise, the actual score is used.
The values of $\alpha$, $\beta$, and $\theta$ were selected intuitively. There is room for empirical studies in the future to determine how people  judge the evaluation quality for different values of $\alpha$, $\beta$, and $\theta$. 

\setlength{\parskip}{1em}

\noindent \textbf{Embedding-based versus F1-measure evaluation:} In Table \ref{table: 4}, we used several sets of predicted keyphrases and gold keyphrases to compare the exact-match F1-measure with greedy matching (no alpha-beta and no threshold), and our proposed embedding-based metric. The last five examples are fabricated pairs of gold keyphrases and predictions to serve as case studies for comparison. For the embedding-based metric, we used both GloVe embeddings pre-trained on general Twitter (General Embd), and GloVe embeddings that we trained on a crisis-related Twitter corpus (Crisis Embd). The crisis-related embeddings were trained on three previously published crisis datasets, specifically, CrisisLexT6 \cite{OlteanuCDV14}, CrisisLexT26 \cite{Olteanu2015} and 2CTweets \cite{schulz2017semantic}, together with tweets collected using the Twitter Streaming API during Hurricanes Harvey, Irma, and Maria, and Mexico Earthquake. The total corpus contains approximately 5.8 million tweets.
As can be seen in Table \ref{table: 4}, the embedding-based scores appear to be more appropriate than the F1 scores, based on a human understanding of the keyphrases. We can also notice (see example 3) that the general Twitter GloVe embeddings can be better for scoring the semantic similarity between general words than the crisis embeddings. This may be because the general Twitter embeddings were trained on a much larger corpus, which is more diverse and general. However, crisis embeddings appear to be better for scoring the similarity between disaster-related words, especially if disaster-related named-entities are involved (see example 4). This is probably because the crisis embeddings were trained on more recent Twitter data and, specifically, on disaster-related data. Not so surprisingly, we also find that crisis embedding-based scores are higher on the disaster data, and general Twitter embedding-based scores are higher on the general data in Table \ref{table:6}. Alpha-beta modulated embedding scores are generally more stringent than embedding scores without the modulation. Without alpha-beta modulation and thresholding, the embedding metrics work as standard greedy matching. The difference made by alpha-beta modulation is most sharply noticeable in examples 6 and 7 where there is a significant difference in the number of gold keyphrases and predicted keyphrases. As can be seen, without alpha-beta modulation, the embedding scores are very high despite the difference in length between the two sets of keyphrases.
\setlength{\parskip}{0em}

\setlength{\textfloatsep}{0em}
\setlength{\floatsep}{0em}
\begin{table}[t]
\small
\centering
\def\arraystretch{1.2}
\begin{tabular}{ | l | c | c | l | } 
\hline
\textbf{Model} & \textbf{P} & \textbf{R} & $\mathbf{F_1}$ \\
\hline
\multicolumn{4}{|l|}{Trained and tested on general data}\\
\hline
J-Bi-LSTM & 83.95\% & 80.81\% & 82.35\%\\
J-Bi-LSTM+ELMo & 84.39\% & \textbf{81.65\%} &  82.99\%\\ 
J-Bi-LSTM+IPA+POS & 84.94\% & 81.63\% & 83.25\%\\
J-Bi-LSTM+ELMo+IPA+POS &  \textbf{85.25\%} & 81.53\% & \textbf{83.35\%}\\
\hline
\multicolumn{4}{|l|}{Trained on general data; tested on disaster data}\\
\hline
J-Bi-LSTM & \textbf{37.49\%} & \textbf{09.92\%} & \textbf{15.68\%}\\
J-Bi-LSTM+ELMo &  35.88\% & 09.69\% & 15.26\%\\ 
J-Bi-LSTM+IPA+POS & 34.53\% & 08.79\% & 14.01\%\\
J-Bi-LSTM+ELMo+IPA+POS &  36.50\% & 09.06\% & 14.51\%\\
\hline
\multicolumn{4}{|l|}{Trained and tested on disaster data}\\
\hline
J-Bi-LSTM & 65.21\% & 61.46\% & 63.28\%\\
J-Bi-LSTM+ELMo & 67.67\% & 59.60\% & 63.38\% \\
J-Bi-LSTM+IPA+POS & \textbf{69.06\%} & \textbf{61.98\%} & \textbf{65.33\%}\\
J-Bi-LSTM+ELMo+IPA+POS & 67.14\% & 60.37\% & 63.57\%\\
\hline
\end{tabular}
\vspace{+.5em}
\caption{Precision, Recall, and F1 scores based on the sequence-labeling evaluation (using exact matches).}
\label{table:5}
\end{table}

\setlength{\textfloatsep}{0em}
\begin{table}[t]
\small
\centering
\def\arraystretch{1.2}
\begin{tabular}{ | l | c | p{3.3em} | p{3em} | } 
\hline
\textbf{Model} & \textbf{F1} & \textbf{General\newline Embd} & \textbf{Crisis\newline Embd}\\
\hline
\multicolumn{4}{|l|}{Trained and tested on general data}\\
\hline
J-Bi-LSTM & 81.31\% & 89.22\% & 86.33\%\\
J-Bi-LSTM+ELMo &  82.24\% & \textbf{90.26\%} & 87.37\%\\ 
J-Bi-LSTM+IPA+POS & 82.11\% & 89.95\% & \textbf{88.84\%}\\
J-Bi-LSTM+ELMo+IPA+POS &  \textbf{82.28\%} & 90.14\% & 87.43\%\\
\hline
\multicolumn{4}{|l|}{Trained on general data; tested on disaster data}\\
\hline
J-Bi-LSTM & \textbf{17.73\%} & 28.30\% & 38.99\%\\
J-Bi-LSTM+ELMo &  17.30\% & \textbf{29.22\%} & \textbf{40.95\%}\\ 
J-Bi-LSTM+IPA+POS & 16.22\% & 29.15\% & 39.47\%\\
J-Bi-LSTM+ELMo+IPA+POS & 16.88\% & 28.58\% & 38.09\%\\
\hline
\multicolumn{4}{|l|}{Trained and tested on disaster data}\\
\hline
J-Bi-LSTM & 63.17\% & 76.97\% & 80.89\%\\
J-Bi-LSTM+ELMo & 62.61\% & 76.55\% & 80.56\%\\
J-Bi-LSTM+IPA+POS &  \textbf{65.32\%} & \textbf{77.30\%} & \textbf{81.09\%}\\
J-Bi-LSTM+ELMo+IPA+POS &  61.33\% & 75.32\% & 79.29\%\\
\hline
\end{tabular}
\vspace{+.5em}
\caption{general and crisis embedding-based scores versus F1 scores, when comparing predicted with gold keyphrases.} 
\label{table:6}
\end{table}

\subsection{Results} 
\setlength{\parskip}{0.3em}

The results of our experiments for the three settings discussed in Section \ref{setting} are presented in Tables \ref{table:5} and \ref{table:6}. Table \ref{table:5} shows precision, recall, and F1 scores, based on exact keyphrase matches (where the order of the keyphrases and the position are both taken into account). Table \ref{table:6} shows F1 scores versus  $GM_{ext}$ embedding-based scores, computed by comparing a set of predicted keyphrases with a set of gold keyphrases (where the order of the keyphrases and the position are both ignored, and duplicate keyphrases are removed).  Thus, while the F1 scores in Tables \ref{table:5} and \ref{table:6} may appear to be inconsistent, we would like to point out that the differences are due to the way the scores are calculated in each table, as explained above. We calculated the F1-scores in Table \ref{table:6} by comparing sets, in order to make them more comparable to the embedding-based scores. 
We will next analyze the performance of each model individually. 
\setlength{\parskip}{1em}

\noindent \textbf{Baseline (J-Bi-LSTM): } The baseline achieved scores in the low-to-mid 80\% range when trained and tested on general data, and scores in the low-to-mid 60\% range when trained and tested on disaster data. Surprisingly, its performance is close  to the performance of other models, which use richer information. Furthermore, the J-Bi-LSTM model produced higher recall than models with ELMo on disaster test data. 
Also, interestingly, the baseline model achieved the best F1 score in Table \ref{table:5}, when trained on general data and tested on disaster data. However, it was not the best scorer if we consider its embedding-based scores in Table \ref{table:6} instead. 

\noindent \textbf{J-Bi-LSTM+ELMo: } The model enhanced with contextual word embeddings achieved a slightly higher score than the baseline when trained and tested on general Twitter data. 
An even better score may have been possible if the ELMo embeddings were pre-trained on a Twitter corpus. However, the F1 performance of this model on disaster data after being trained on general data is lower than the performance of the baseline, but its embedding-based scores are higher. There is no significant difference between the overall F1 score of the baseline and the F1 score of this model when trained and tested on disaster data. Although the model seems to get slightly better precision, it struggles on recall more than the baseline. 

\noindent \textbf{J-Bi-LSTM+IPA+POS: } The addition of POS, IPA and phonological feature information generally improves the performance of the J-Bi-LSTM+IPA+POS model over the baseline. The model even gets a better score than the J-Bi-LSTM+ELMo model, and it achieves similar performance to J-Bi-LSTM+ELMo+IPA+POS on the general test data. So, simply adding POS, IPA and phonological information may eliminate the need for ELMo embeddings. This could be because both ELMo and IPA (with phonological features) can model some similar character-level information (like morphology). Furthermore, as mentioned above, ELMo may not be expressing its full potential in the Twitter domain because it was not pre-trained on a noisy social media corpus. 
Interestingly, the J-Bi-LSTM+IPA+POS appears to have the worst F1 score when trained on general data and tested on disaster data, but it performs the best in terms of the crisis embedding-based metric. When trained and tested on disaster data, this model is also one of the better performers. 

\noindent \textbf{J-Bi-LSTM+ELMo+IPA+POS: } Finally, the model enhanced with ELMo embeddings, and also POS, IPA and phonological feature information, achieved the best overall F1 score when trained and tested on general data, but not in the other two scenarios. It seems that the ELMo embeddings and the POS/phonetics may contain somewhat overlapping information. ELMo can capture some character level information, which may be why adding further character level information from IPA and phonological features do not give a significant boost in performance. Overall the performance of this model is similar to the performance of the JRNN+ELMo model on the disaster data.  We also considered training models using crisis embeddings 
 but they did not help.

 \vspace{-2mm}
\noindent In the remaining of this section, we will use the results in Tables \ref{table:5} and \ref{table:6} to answer our original questions. First, regarding question (1), we can say that the models enhanced with contextual and/or POS/phenomes information generally perform better than the baseline model, although the baseline is not much worse. 
\setlength{\parskip}{0em}

\begin{figure*}[t]
  \centering
  \subfloat[Disaster training data]
  {\includegraphics[width=28em]{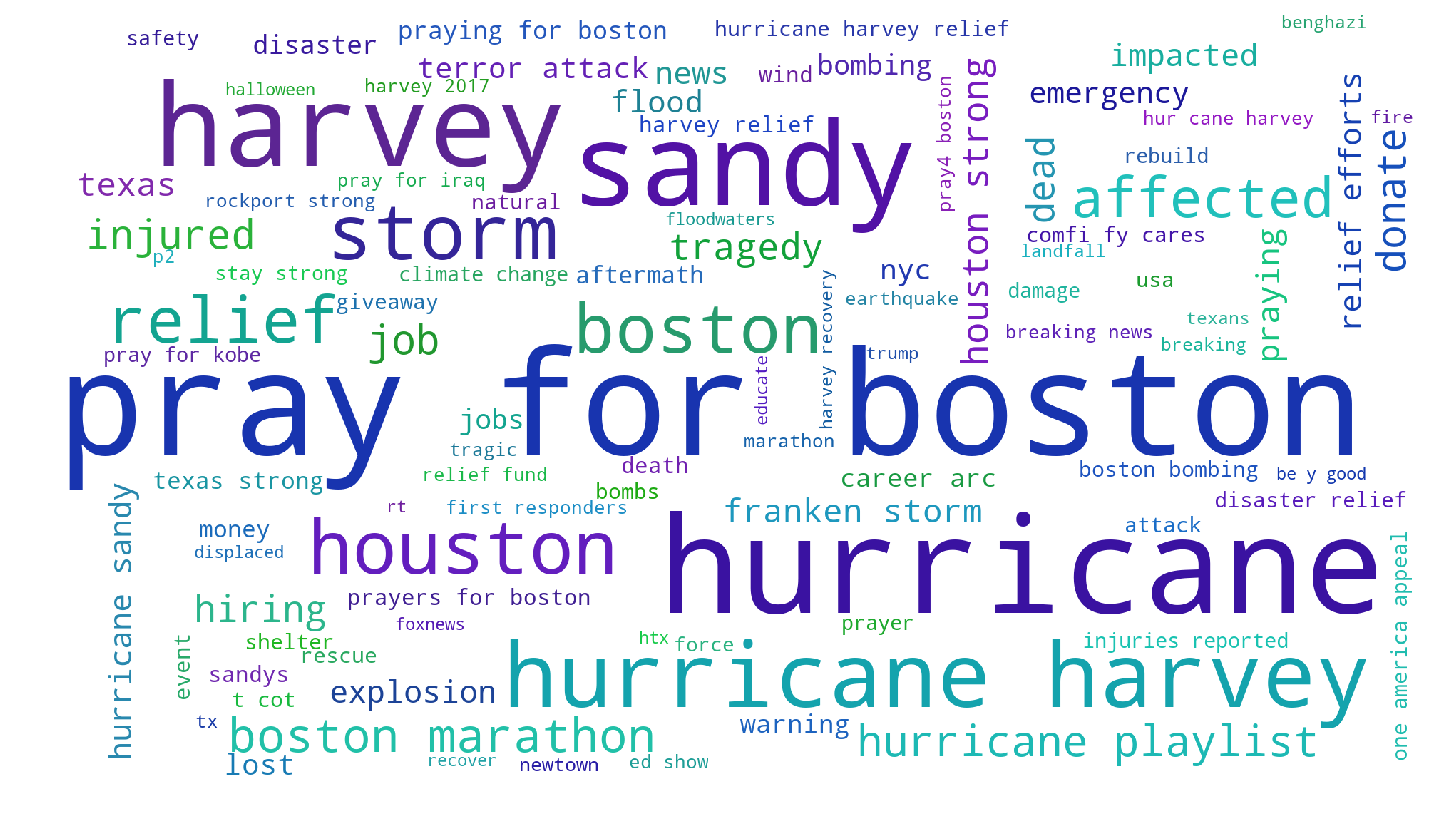}}
  \centering
  \subfloat[Disaster test data]
  {\includegraphics[width=28em]{disaster_gold.png}}
  \caption{Word clouds of the top 100 most frequent gold keyphrases from disaster training data (a) and disaster test data (b).}
  \label{fig:3}
\end{figure*}

\begin{figure*}[t]
  \centering
  \subfloat[Predictions by model trained on general data]
  {\includegraphics[width=28em]{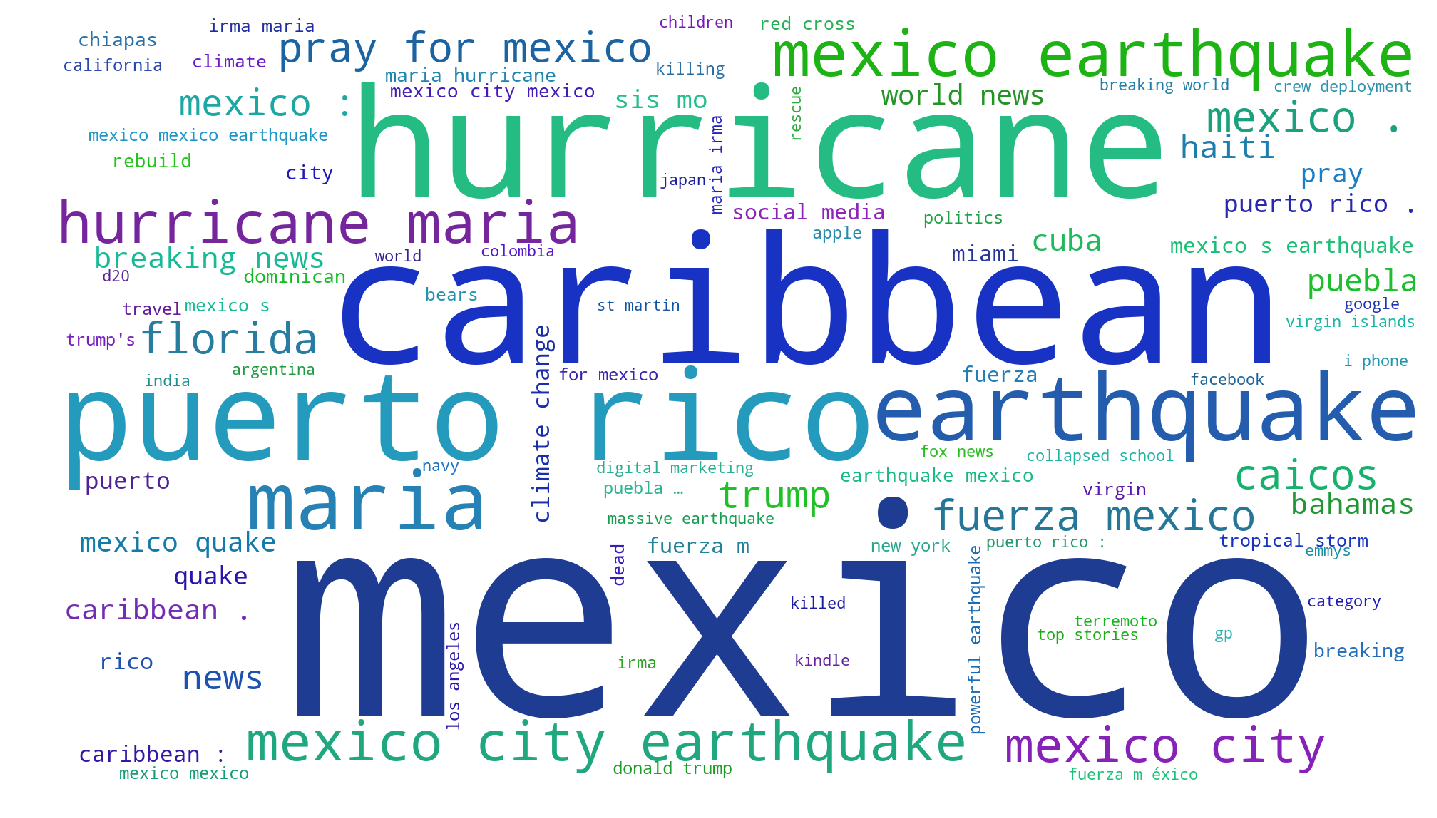}}
  \centering
  \subfloat[Predictions by model trained on disaster data]
  {\includegraphics[width=28em]{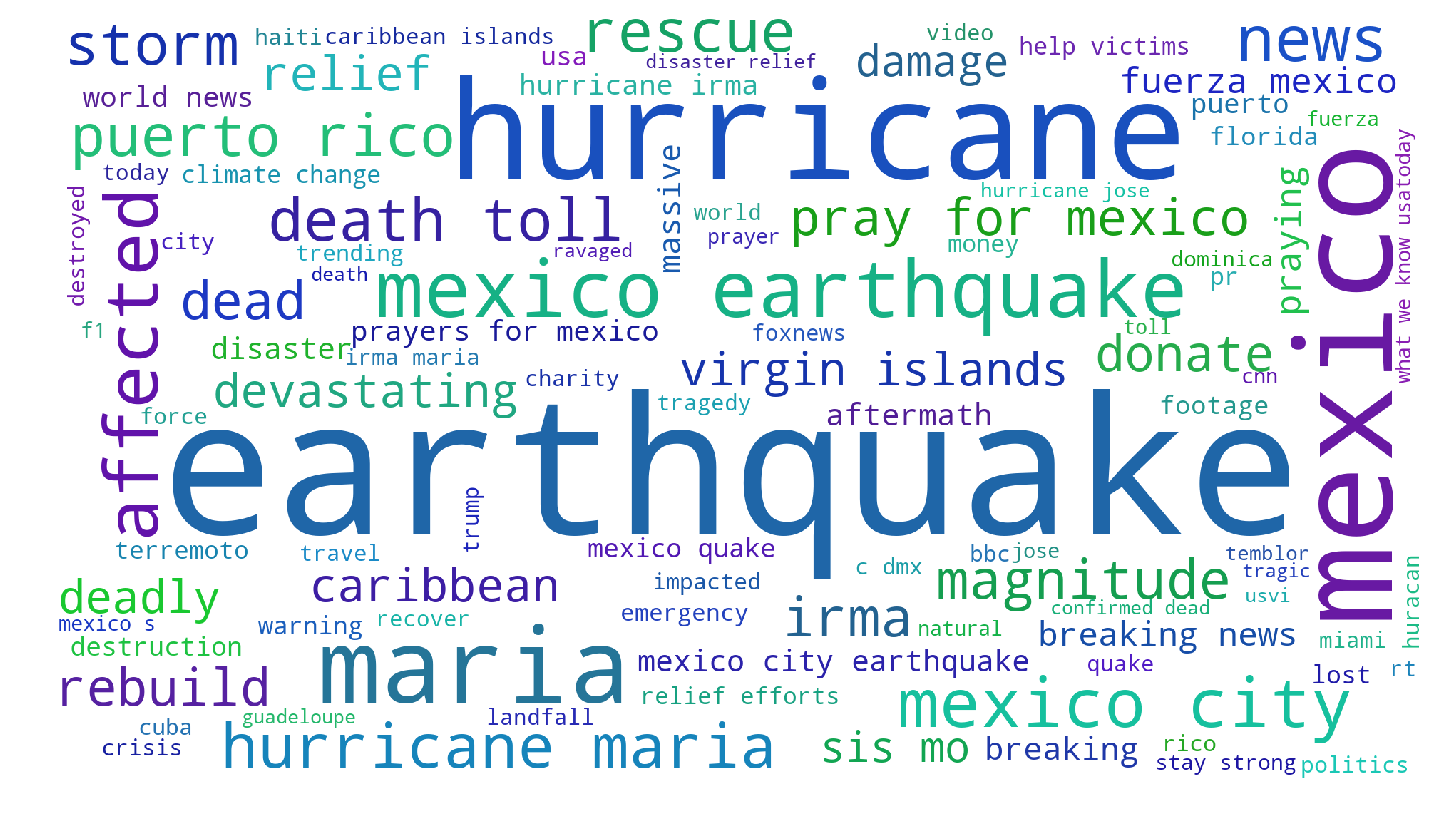}}
  \caption{Word clouds of the top 100 most frequent keyphrases from the disaster test data, as predicted by the J-Bi-LSTM+IPA+POS model trained on the general data (a) and J-Bi-LSTM+IPA+POS trained on disaster data (b).}
  \label{fig:4}
\end{figure*}

Regarding our original question (2), we noted that the models achieved much lower F1 scores when trained on general data and tested on disaster data. It should be expected that the domain differences contribute to the low score. However, there is another important reason. As we mentioned before in Section 4, the average number of keyphrases per disaster tweet is approximately twice the average number of keyphrases per general tweet. Thus, it is likely that the models, when trained on the general tweet data, learn to predict one keyphrase per sample on average. This results in a substantial hit to their recall when tested on the disaster data, which presents more variety in the number of keyphrases per tweet. Nevertheless, as can be seen from figure \ref{fig:4}(a), a model trained on general data, can still predict keyphrases similar to the gold (Figure \ref{fig:3}(b)). Furthermore, their embedding-based scores are also much higher than their overall F1 scores. 
Due to the fundamental differences in the two datasets (as discussed in more details in Section 4), a model may learn some parameters which are useful for general tweets, but the same parameters may be counter-productive in the case of tweets from the disaster test data. This may be why the models that performed better on general data struggled more to generalize on disaster data. But the results are not too straightforward, because some of those models did generalize better on disaster data when considering their embedding-based scores. Overall, it may be better to be cautious about judging the models' capabilities based on their performance on a domain they were not even trained on. 

Finally, to answer question (3), we can claim that the models trained and tested on the disaster data can perform significantly better than the models trained on general data and tested on disaster data. However, the performance scores on the disaster test data are overall lower than those on the general test data. This is because we used a rather challenging test set for the disaster data, as discussed earlier. Furthermore, the low diversity of keyphrases in the disaster training data can encourage the models to memorize more (and thus overfit) instead of learning more general abstract features. Nevertheless, they can still predict keyphrases that are fairly similar to the gold (see Figures \ref{fig:3}(b) and \ref{fig:4}(b)), and their embedding-based scores are better than the  F1 scores. 
\setlength{\parskip}{0em}

\setlength{\textfloatsep}{2em}

\vspace{-1mm}
\section{Conclusions and Future Work}
We showed that the Joint-layer Bi-LSTM model can be a promising approach for extracting disaster-related keyphrases from Tweets. We also showed that ELMo embeddings and/or phonetics and phonological features, along with POS-tags, can generally improve the performance of the model. 
We discussed how the exact match F1 can be an unsuitable metric for evaluating the quality of the predicted keyphrases, and we showed that embedding-based metrics are more appropriate for this task. We further introduced novel extensions to the embedding-based metrics, to account for differences in terms of the number of predicted and gold keyphrases. 

In future work, we would like to focus on abstractive keyphrase generation \cite{meng2017deep}. Taking a generative approach as opposed to an extractive one would allow us to generate keyphrases that are absent from the source text but still appropriate given the semantics of the text. This can further enhance keyphrase-based search of tweets related to a particular context (including disaster context). 

\section{Errata}

There was a bug in calculating set-based evaluation measures (Table 4) which skipped data with no predictions. We fixed it and updated the results. The updated results are still consistent with our statements. 

\newpage

\section*{Acknowledgements}
 We thank the National Science Foundation (NSF) and Amazon Web Services for support from grants IIS-1741345 and IIS-1912887, which supported the research and the computation in this study.  We also thank NSF for support from the grants IIS-1526542, IIS-1423337, IIS-1652674, and CMMI-1541155. 

\bibliographystyle{ACM-Reference-Format}
\raggedright
\bibliography{main}

\end{document}